# Prediction of Self-Assembled Dewetted Nanostructures for Photonics Applications via a Continuum Mechanics Framework

L. Martin-Monier[1,*], P.G. Ledda[2,*], P.L Piveteau[1], F. Gallaire[2,†], F. Sorin[1,†]

[1] Laboratory of Photonic Materials and Fiber Devices, IMX, École Polytechnique Fédérale de Lausanne, 1015, Lausanne, Switzerland.

[2] Laboratory of Fluid Mechanics and Instabilities, IGM, École Polytechnique Fédérale de Lausanne, 1015, Lausanne, Switzerland.

*: These authors contributed equally to this work

†: Corresponding Author

## Abstract

When a liquid film lies on a non-wettable substrate, the configuration is unstable and the film then retracts from a solid substrate to form droplets. This phenomenon, known as dewetting, commonly leads to undesirable morphological changes. Nevertheless, recent works have demonstrated the possibility to harness dewetting by employing templated substrates with a degree of precision on par with advanced lithographic processes for high-performance nanophotonic applications. Since resonant behavior is highly sensitive to geometrical changes, predicting quantitatively dewetting dynamics is of high interest. In this work, we develop a continuum model that predicts the evolution of a thin film on a patterned substrate, from the initial reflow to the nucleation and growth of holes. We provide an operative framework based on macroscopic measurements to model the intermolecular interactions at the origin of the dewetting process, involving length scales that span from sub-nm to μm. A comparison of experimental and simulated results shows that the model can accurately predict the final distributions, thereby offering novel predictive tools to tailor the optical response of dewetted nanostructures.

**Introduction**

Flows of thin films over substrates are of central interest owing to their ubiquity in natural as well as industrial environments. Depending on the interaction between film, substrate and surrounding environment, thin films may dewet and re-arrange into droplets.[1] The miniaturization of modern devices involving ultra-thin layers has brought a novel focus on the question of dewetting. This ubiquitous phenomenon threatens the integrity of thin films, typically yielding semi-ordered tessellation patterns. This has often been an undesirable phenomenon, jeopardizing the film morphology with little practical use. Nevertheless, a few research endeavors have investigated the potential of dewetting as an efficient self-assembly process. Several approaches have been proposed based on selective wetting[2,3] or spatially-patterned electric fields.[4,5] These investigations have demonstrated interesting possibilities but remained limited in terms of materials, scalability, geometry or resolution. By combining engineered substrates with functional materials, thermal dewetting has emerged as a viable alternative, as illustrated in FIG. 1.[6,7,8,9] In opposition to mono- or polycrystalline thin films that dewet according to surface diffusion mechanisms, amorphous thin films dewet following bulky viscous flow mechanisms. While surface diffusion over templates has been the object of several investigations,[6,10,11] the flow of amorphous films over pre-patterned substrates have thus far only made the object of linear stability analyses, which fail to predict the complete dewetting dynamics.[12,13] This is surprising, as isotropic material properties associated with amorphous materials allow for improved control of the re-arrangement mechanism and higher complexity in the final microstructures. Resorting to high-index dielectric glasses bears relevance in nanophotonics,[14] where the control of the nano-resonator geometry allows for wide engineering of scattering and resonant properties. To increase the method's accuracy and further extend its realm of application, the prediction in the most precise manner of the final microstructure based on the exact substrate shape and film-substrate interactions is of paramount importance, which motivates the present study.

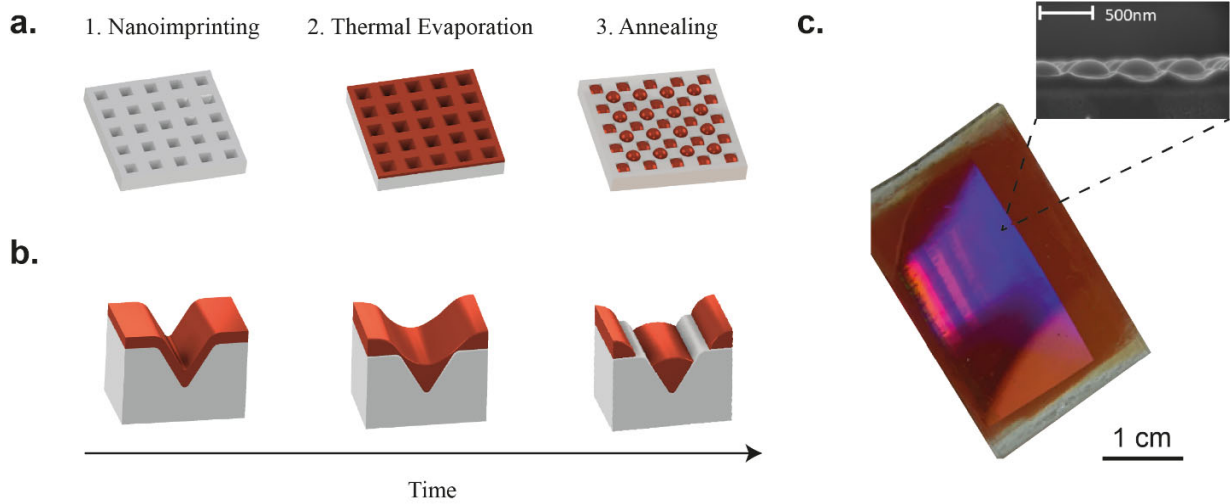

FIG. 1 – Description of processing scheme. (a) Schematic describing the three main process steps: 1. Nanoimprinting of a nanoscale texture onto a sol-gel or UV curable substrate from a silicon master mold; 2. Thermal evaporation of a thin (<100 nm) optical glass layer; 3. Annealing to induce ordered re-arrangement of the film according to the underlying texture. (b) Time evolution of the thin glassy layer under annealing. (c) Optical photograph of a 350 nm meta-array of selenium nanoparticles.

Herein, we propose a continuum model predicting the evolution of a templated film evolving over pre-patterned substrates, based on the modeling of intermolecular interactions occurring on various substrates, for contact angles less than 90°. By comparing experimental and simulated thickness profiles over various patterns, we show that the proposed model is suitable for the accurate prediction of the final morphology, and over several length scales (from nm to $\mu$m scale). We thereby offer improved control of the dewetting patterns, allowing for the realization of precise architectures relevant to nanophotonics.

**Model Description**

*Dewetting on flat substrates: validation*

An accurate dynamic description of dewetting constitutes a particularly challenging problem. The theoretical framework for the description of fluid flow is based on approaches from continuum mechanics. We initially consider a flat horizontal substrate and introduce a coordinate system $(x, y, z)$, where the z direction, along which the film thickness is measured, coincides with the vertical one, as sketched in FIG. 2(a). The integration of the Navier-Stokes equations along the z direction, under the classical assumptions of the long-wavelength approximation,[15,16] leads to an evolution equation for the thin film thickness h in the (x, y) directions, so-called lubrication or thin film equation:[16]

$$\frac{\partial h}{\partial t} = -\frac{1}{3\mu} \nabla \cdot \left( h^3 \nabla (\gamma \nabla \varkappa - \Pi(h)) \right) \tag{1}$$

Where $\kappa$ is the curvature, $\nabla$ operates in the (x, y) plane, $\mu$ is the fluid dynamic viscosity, $\gamma$ is the surface tension coefficient between the fluid and the air, and $\Pi$ is the so-called disjoining pressure. In the classical framework of the long-wavelength approximation, the curvature is implemented with its linearized version, i.e. $\varkappa = \nabla^2 h$, that holds for small slopes of the film. As discussed further on, we resort here to the complete curvature to properly account for arbitrary height profiles. Note that, since the thickness is a single-valued function of the position, contact angles greater than 90° cannot be described by this model.

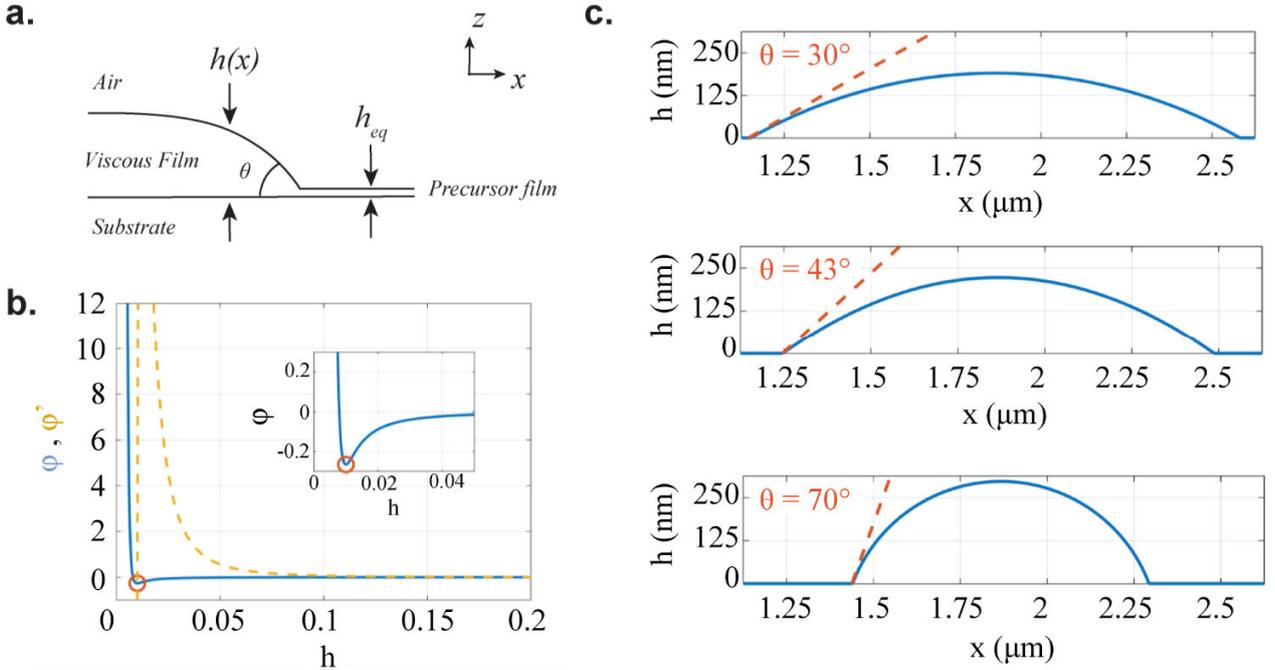

FIG. 2 – Precursor film approach implemented on flat substrates. (a) Schematic describing the components of the system and the associated parameters in the precursor film approach. (b) Lennard-Jones potential associated with the viscous film on a flat substrate. The potential (blue curve) shows a minimum, which is further indicated by its derivative (dashed yellow curve) and by the zoom in the inset graph. The minimum of the potential corresponds to a film thickness that coincides with the precursor film thickness $h_{eq}$ indicated in (a). (c) Comparison between the results of the thin film model (blue solid lines) and the prediction (orange dashed lines) given by the two-dimensional equation (4), for $h_i = 25$ nm and different contact angles reported in the figures. From the top to the bottom: A = $5.10^{-20}$ J; $1.10^{-19}$ J; $2.5.10^{-19}$ J.

The disjoining pressure term is assumed to stem from a classical Lennard-Jones type potential:[17,18,19]

$$\varphi(h) = \frac{B}{h^8} - \frac{A}{12\pi h^2} \qquad (2)$$

where $A=A_{123}$ is the so-called Hamaker constant of the system substrate (1) - film (2) - air (3) and B is the Born coefficient, employed to model respectively the molecular long-range attractive and short-range repulsive forces. The combination of a repulsive and an attractive term defines a minimum of the potential for an equilibrium "precursor" film thickness $h_{eq} = (48\pi B/A)^{1/6}$, obtained by imposing $\varphi'(h_{eq}) = 0$ (FIG. 2 (b)).

The force derived from the Lennard-Jones potential stems from an imbalance in the interactions between the various constituent molecules. This imbalance is classically embedded in the Hamaker constant $A_{123}$, which establishes the influence of constituent materials in long-range interactions, in the presence of multiple bodies according to Lifschitz theory.[20,21,22,23] The previously introduced Lennard-Jones potential is linked to the disjoining pressure $\Pi$ through:

$$\Pi = -\frac{\partial \varphi}{\partial h} = \frac{8B}{h^9} - \frac{A}{6\pi h^3} \qquad (3)$$

A positive Hamaker constant induces destabilizing pressure gradients for films larger than the equilibrium thickness $h_{eq}$. When a region of the film reaches the precursor film thickness $h_{eq}$, the local equilibrium at the interface between the precursor film and the thicker regions defines an apparent contact angle $\theta$ (FIG. 2(a)) given by:[17,24]

$$1 + \tan^2 \theta = \left(\frac{\varphi(h_{eq})}{\gamma} + 1\right)^{-2} \qquad (4)$$

Considering solely angles between 0° to 90°, relation (4) provides a bijective relationship between the contact angle and the precursor film $h_{eq}$.

To validate this approach, we now proceed to simulate the evolution of a thin film and evaluate the resulting contact angle. The thin film equation is implemented with the full expression of the interface curvature:[25,26,27,28]

$$\varkappa_f = -\vec{\nabla} \cdot n \tag{5}$$

$$n = \frac{1}{\left(1 + (\frac{\partial h}{\partial x})^2 + (\frac{\partial h}{\partial y})^2\right)^{1/2}} \cdot \begin{bmatrix} -\frac{\partial h}{\partial x} \\ -\frac{\partial h}{\partial y} \end{bmatrix} \tag{6}$$

where n embeds the x and y component of the normal of the fluid free surface; the problem is completed with the disjoining pressure $\Pi(h)$ detailed above. The model with the full expression of the curvature, despite its simplicity, showed a very good agreement with various experimental measurements, even for cases in which the typical assumptions of the long wave approximation are not respected[26,27,28]. To verify the consistency of the relation equilibrium thickness-contact angle, we perform numerical simulations with the finite-element solver COMSOL Multiphysics by implementing the weak form of equations (1), (3), (5) and (6) in conservative form (see Methods). We choose three different values of the contact angles and three different values of the Hamaker constant, and determine the corresponding values of the Born coefficient B. We then determine the Born coefficient by using relations (2), (3) and (4). Simulation results with initial thickness $h_i = 25$ nm and $\gamma = 3.10^{-2}$ N.m$^{-1}$ are shown in FIG. 2(c), for three different values of A = $5.10^{-20}$ J, $1.10^{-19}$ J, and $2.5.10^{-19}$ J. The numerical values of the contact angle matched the predicted ones with an accuracy below 1°, which validates the proposed approach on flat substrates. The implementation of the complete curvature is essential to yield the proper results. The linearized curvature $\varkappa_f = \nabla^2 h$ in the long wavelength approximation gives contact angle values with over 10° error compared to the target value. Despite the small size of the final drop states involved, the importance of the complete curvature expression of the curvature to recover the final static shape is remarkable, as already highlighted.[26]

It is also important to note that the presence of a precursor film implies a loss of volume proportional to the precursor film thickness. The volume error associated with this choice is in all cases presented here inferior to 1%, and thus neglected.

At this stage, an important question arises with the choice of the contact angle. During dewetting, it is common that contact angles evolve dynamically owing to the elasticity of the triple line[1]. In this framework, it is observed that the final contact angle does not show a strong dispersion in the final stage of dewetting (inferior to 6°, see FIG. SI 1). Hysteresis is thus neglected in the rest of this work. Despite this assumption, our model still allows for a relatively accurate prediction of experimental observations, as discussed below.

In the rare previous works that developed a model based on the Lennard-Jones potential,[16,17,19] the contact angle was inferred from accurate Hamaker constant and Born coefficient data, with a good agreement between theoretically derived and experimentally measured contact angles. This approach assumes the prior knowledge of the Born coefficient, significantly harder to quantify than the Hamaker constant, and thus constitutes a significant limitation for broader use of such modeling scheme. Moreover, this requires the knowledge of the precursor film thickness, a challenging quantity to experimentally measure (typically in the Ångström range). In this work, we propose to use the Born coefficient, and thus the equilibrium thickness given by (4), as independent parameter to match the experimental and modeled contact angles in analogy to the procedure in FIG. 2(c) (see Methods). In contrast with previous works, the knowledge of the equilibrium contact angle, Hamaker constant, and surface tension coefficient is sufficient to effectively model the thin film dynamics.

*Templated Dewetting*

We now turn to the evolution of a film with fixed contact angle $\theta$ over a pre-patterned (or "templated") substrate, of height profile $h_s$ measured along the z direction starting from the horizontal reference previously introduced (see FIG. 3(a)). In this work, we consider two different types of templates, (i) lines or two-dimensional templates, characterized by triangular trenches of base W and period P along the $x$ direction, invariant along the direction orthogonal to the periodicity one ($y$ direction), and (ii) pyramids or three-dimensional templates, characterized by pyramidal trenches with periodicity P along both $x$ and $y$ directions. We resort to amorphous $As_2Se_3$ thin films thermally evaporated on textured UV-curable polymers or silica substrates (See Methods section for further detail regarding the materials employed). For templates shown in FIG. 3(a), the fabrication process[14] leads to dewetted patterns invariant along the y-direction.

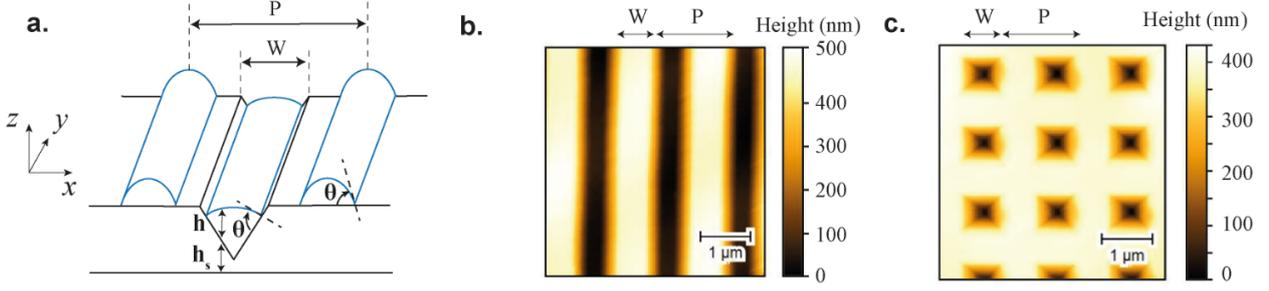

FIG. 3 – Two-dimensional and three-dimensional templates. (a) Schematic introducing the contact angle θ, the film height profile h, the substrate height profile $h_s$, the inverted pyramid base width W, and the template period P. (b)-(c) Atomic Force Microscopy images of textures nanoimprinted on an Ormocomp® substrate: (b) triangular grooves (two-dimensional template), and square arrays of inverted pyramids (three-dimensional template).

To accurately predict the thin film dynamics and the resulting microstructure, the thin film equation must be adapted to account for the role of the underlying substrate. The total surface curvature $\varkappa$ in this new configuration is now given by the curvature of the total elevation of the free surface $h + h_s$:[29]

$$\varkappa = -\vec{\nabla} \cdot \mathbf{n_t} \tag{7}$$

$$\mathbf{n_t} = \frac{1}{\left(1 + (\frac{\partial h}{\partial x} + \frac{\partial h_s}{\partial x})^2 + (\frac{\partial h}{\partial y} + \frac{\partial h_s}{\partial y})^2\right)^{1/2}} \cdot \begin{bmatrix} \left(-\frac{\partial h}{\partial x} - \frac{\partial h_s}{\partial x}\right) \\ \left(-\frac{\partial h}{\partial y} - \frac{\partial h_s}{\partial y}\right) \end{bmatrix} \tag{8}$$

Another difference with the flat substrate case lies in the definition of the film thickness in the Lennard-Jones potential. Recalling the definition of the potential (Equation (2)), the contact angle depends on the equilibrium thickness. A proper definition of the thickness is therefore crucial to reproduce identical contact angles over the whole substrate. On a flat substrate, thickness is defined straightforwardly as a vertical projection. However, in the general case with an underlying substrate, the accurate film thickness is defined as the shortest distance between film-air interface and film-substrate, i.e. the projection given by:

$$h^* = h \cos\left(\operatorname{atan}\left(\left(\frac{\partial h_s}{\partial x}\right)^2 + \left(\frac{\partial h_s}{\partial y}\right)^2\right)^{1/2}\right) \tag{9}$$

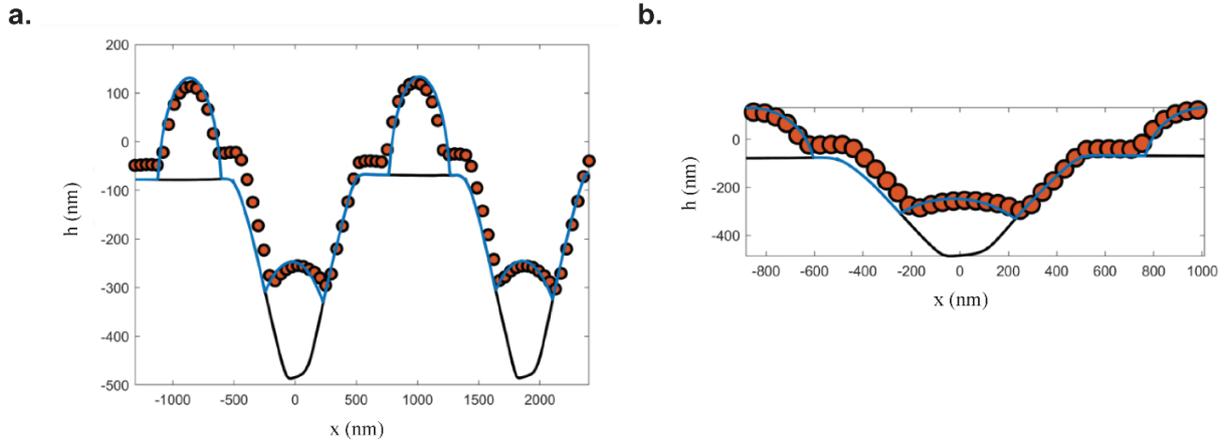

FIG. 4 – Simulated vs. experimental dewetted profiles in the 2 D case. (a)Simulated dewetted profile (blue) and experimental AFM dewetted profile (orange dots) in the case of a 2 μm period line pattern with a 850 nm pyramid base and a 60 nm initial film thickness of $As_2Se_3$. (b) Close-in view of (a) on a single period, showing the match between experimental and simulated contact angles.

We adapt the numerical scheme with the latter thickness definition and ensure that it yields the desired contact angle over substrates with arbitrary inclination (See FIG. 4).

To validate the proposed scheme, numerical simulations using experimental atomic scanning microscope profiles of nanoimprinted substrates are calculated. We take as initial condition a constant flat thickness that matches the imposed thicknesses in the experimental campaign (See Appendix). The experimental film profile upon dewetting is then superposed to compare the match between experimental and simulated data. First investigating the two-dimensional case, the experimental and simulated dewetted film profile is compared in FIG. 4 (period 2 μm, inter-pyramid spacing 1.1 μm, film thickness 60 nm, contact angle θ = 85°). Additional comparative results in the two-dimensional case are provided in the supplementary material (see FIG. SI 2). The numerical scheme is further validated in three dimensions, using a pyramid with largely reduced spacing (inter-pyramid spacing 150 nm, period 1.1 μm, thickness 60 nm, see FIG. 5). Remarkably, the model reproduces with accuracy the experimental height profile over the range of thicknesses considered in this work. It is interesting to note that the proposed framework also predicts a thickness threshold $h_{crit}$ above which the final film equilibrium upon simulation leads to a flat film covering the full substrate, instead of isolated droplets, as shown in FIG. SI 3 for an initial thickness of $h_i$=100 nm. This ultimately leads to dewetting according to nucleation and growth holes with random location instead of a well-prescribed location. Interestingly, the random nucleation and growth of holes are observed experimentally at around 80 nm (contact angle 64.5°), which is well in line with the results of FIG. SI 3. These results highlight the relevance of continuum mechanics schemes even at thicknesses that become comparable with atomistic length scales.

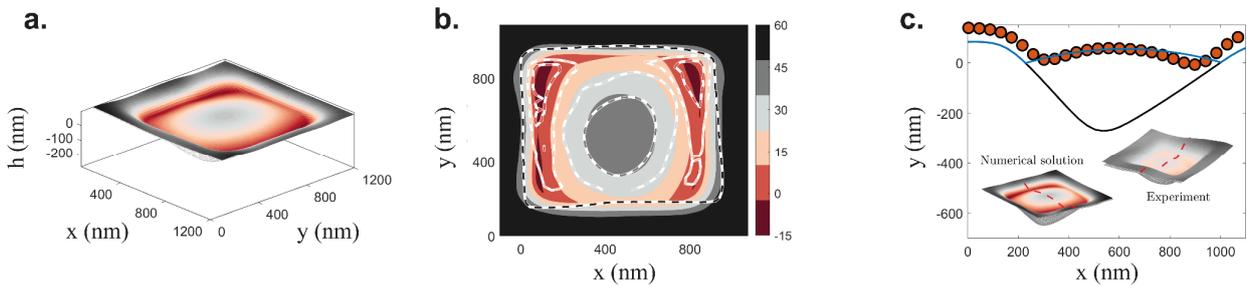

FIG. 5 – Simulated vs. experimental dewetted profiles in the three-dimensional case, in the case of a 1.1 μm period inverted pyramid pattern with an 850 nm pyramid base. (a) Three-dimensional visualization of the simulated dewetted profile. (b) Comparison of the numerical (coloured isocontours) and experimental (dashed isocontours) dewetted profiles. (c) Comparison along one section of the simulated dewetted profile (Solid blue surface) with the experimental AFM dewetted profile (orange dots).

Given that the rearrangement of material is fundamentally linked to the increase in local curvature, the influence of curvature amplitude on final structures is further investigated (see FIG. SI 4). Interestingly, simulated transitory states may differ based on pyramid base edge curvature, giving rise for some cases to pinning behavior. Nevertheless, the final dewetted architectures appear independent on pyramid edge curvature for sufficiently long simulation times.

A particularity of dewetting in such pyramid arrays pertains to the distribution of material in the final microstructure, which widely varies depending on the spacing-to-period ratio. In FIG. 6 (a)-(c), the final volume inside the pyramid $V_{in}$ over the total volume $V_{tot}$ is evaluated by simulation in the two-dimensional case. In the case where the final pyramid volume would be solely constituted from the material initially deposited inside the trench, the pyramid volume should be constant with spacing. This would therefore impose a well-defined law, referred in this work as volume conservation, according to which the initial volume inside and outside the periodic trenches is conserved: $\frac{V_{in}}{V_{tot}} = \frac{h_i.W}{h_i.P}$ and $\frac{V_{out}}{V_{tot}} = \frac{h_i.(P-W)}{h_i.P}$. Both the experimental and numerical results show a clear deviation from the volume conservation law at low values of the spacing-to -period- ratio, labeled S, where the film located at the pyramid edges is preferentially pulled inside the pyramid (see FIG. 6 (c)-(d)). On the contrary, at larger spacings to period ratios, the ratio $V_{in}/V_{tot}$ follows closely the volume conservation law. While the thickness dependence does not appear in the experimental volume analysis, the deviating trend is observed for all configurations investigated, including lines with incremental spacing as well as pyramids with incremental spacing along both one and two dimensions. At reduced spacing, the absence of droplets in between pyramids ('mesa') is observed for spacing-to-period ratios (written S) of up to 0.35, while at very large spacing, the instability in the top plane gives rises to a double distribution in size and thus to a new architecture. To provide the reader with an overview of the architectures as a function of spacing and thickness, a diagram is provided in FIG. 6 (e). Additional SEM images showing the full structural transition with the thickness and spacing-to-period ratio are provided in FIG. SI 5.

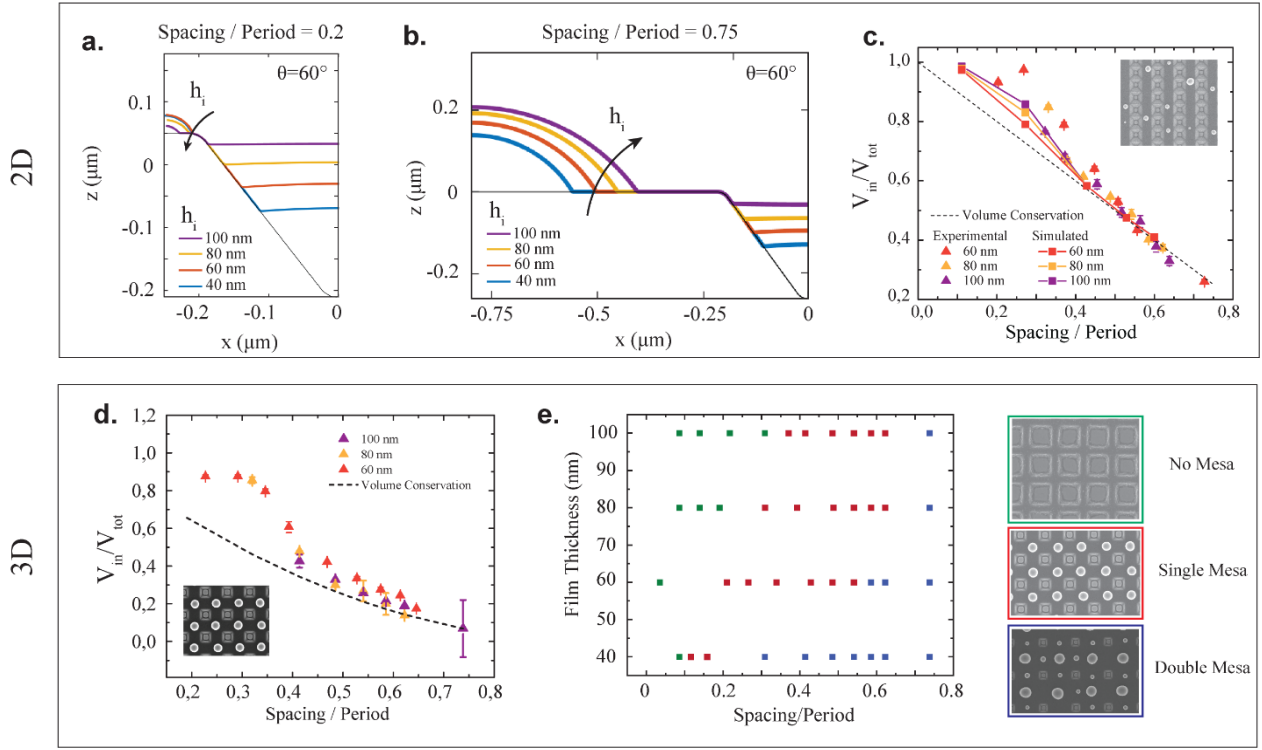

FIG. 6 – Distribution of film material upon dewetting. (a) Two-dimensional final film profiles upon dewetting for spacing over period ratios S of S=0.2 (left) and S=0.75 (right). The mesa particle size grows with film thickness for S=0.75, while the trend is reversed for S=0.2. (b) Simulated pyramid volume over total volume ratio in the case of a two-dimensional geometry. Colors indicate the initial film thickness. Blue : 40 nm, Orange : 60 nm, Yellow : 80 nm, Purple : 100 nm. (c) Experimental (Triangular points) and Simulated (Solid line) pyramid volume over total volume ratio in the case of an inverted pyramid array with increasing distance along a single direction. (d) Experimental pyramid volume over total volume ratio in the case of an inverted pyramid array with increasing spacing along the two principal directions. (e) Structure diagram associated with inverted pyramids with varying spacing to period ratio and film thickness. The film material is composed of $As_2Se_3$ onto a plasma-treated Ormocomp® substrate.

*Application in Photonics: Phase Control*

We now turn to exploit this in-depth understanding and control of template dewetting to realize advanced optical metasurfaces. Ordered high-index nanoparticles bear particular importance for metasurfaces or meta-gratings, which enable to tailor the phase, amplitude and polarization of light over reduced thicknesses, in stark contrast with current bulky optical components.[30,31] By engineering the coupling between the various Mie modes through geometry, recent works have demonstrated the ability to tailor the emitted phase. These so-called Huygens meta-atoms provide control over the phase covering the full 0-2$\pi$ range, hence enabling arbitrarily tailored phase profiles.[32,33,34] Several demonstrations based on this concept have been implemented such as lensing. Nevertheless, achieving full control over phase imposes stringent requirements, since geometrical changes of a few tens of nanometers may have a significant impact on the optical response. Given the high accuracy in terms of both position and spacing in template dewetting, quasi three-dimensional structures present remarkable opportunities in terms of phase modulation.[14]

Let us now focus on the optical properties arising from periodic architectures based on inverted pyramids with increasing spacing along the two principal axes (Single Mesa architecture in FIG. 7). We proceed to evaluate the meta-assembly spectrum in reflection for three distinct period values (P = 1270 nm, P = 1440 nm, P = 1550 nm), keeping the pyramid base constant at 850 nm (see FIG. 7(a)). The experimental spectra of line arrays are further compared with the equivalent simulated shapes. The simulated geometrical shapes rely on both experimental equilibrium contact angles and the volume conservation criterion discussed in FIG. 7(b)-(d), which together define a relationship between evaporated film thickness and line width. Since the system is highly sensitive to slight geometrical changes as low as 10 nm, experimental and simulated reflection spectra appear relatively well in line. We now turn to the evolution in phase for a range of periods ranging from 1200 nm to 1500 nm. Interestingly, interference between the various individual particle Mie mode cumulate to yield a cumulative phase shift over the complete $2\pi$ phase range at $\lambda = 784$ nm. The phase shift is gradual, spanning from 1200 nm to 1500 nm, hence allowing for phase control with experimentally attainable accuracies. One can also note a particularly sharp 2π phase shift occurring for periods around 1410 nm, which highlights the sensitivity of imparted phase on geometrical parameters.

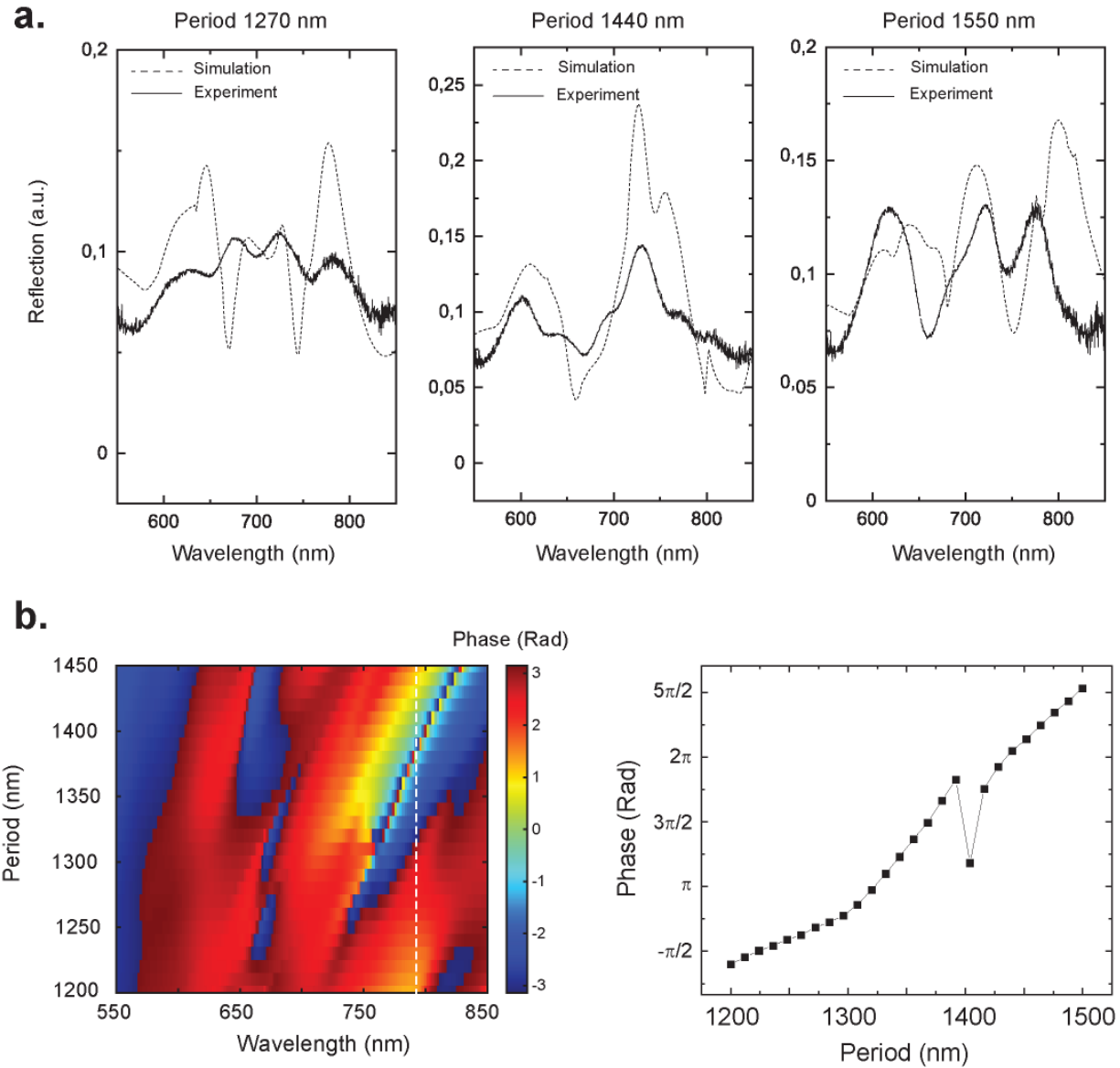

FIG. 7 – Phase control for meta-gratings. (a) Simulated (Dash) and experimental (solid) reflection spectra for three periodicities (P=1270 nm, 1440 nm, 1550 nm) and fixed pyramid size of 850 nm. The spectra are not normalized with their own maximum but compared in absolute value. (b) (left) Colormap representing the phase imparted by the quasi 3D resonating structure to the outbound beam, as a function of the period and wavelength. The pyramid width W is fixed at 850 nm. (right) Plot representing the phase as a function of the period at 784 nm, corresponding to the white dash line reported in (left).

A similar study for quasi-3D line array (see FIG. SI 6) also yields a coherent match between experimental and simulated spectra. Considering the spacing to period ratios studied here, we can resort solely to the volume conservation hypothesis to link thickness with resulting geometrical parameters. This allows for a completely "blind" implementation of geometry in FDTD simulation software, which bears significant advantages in terms of design scalability. The phase profile shows extended phase control possibilities around 684 nm (see FIG. SI 6), with an extended range of periods to tune the phase, spanning from 1400 nm to 1900 nm.

**Conclusion**

To conclude, we have introduced a novel modeling framework for the dewetting of films over a templated substrate based on a precursor film approach. By resorting to a Lennard-Jones potential model, solutions to the dewetting problem have been identified. Comparisons between final simulated and experimental film profiles show quantitative agreement, thereby providing for the first time an accurate predictive model for the fabrication of nanostructures via dewetting on templated substrates. Finally, we demonstrate how this fine understanding of the resulting geometries paves the way for wavefront control in quasi three-dimensional architectures. Further works to accelerate the convergence of the model in 3D would allow for an end-to-end framework that combines three-dimensional dewetting models with photonic simulation tools, enabling to directly simulate optical properties of a dewetted pattern based on simple input parameters (mainly initial thickness and pattern profile). This would thereby significantly expand the opportunities of fabrication of self-assembled nanostructures, with a precision comparable to advanced lithographic processes. These considerations find immediate applications in the context of metasurfaces, and beyond in the field of nanophotonics.

**Acknowledgements**

The authors would like to thank the scientific staff from the Center of Micro and nanotechnology (CMi) and the Center for electron microscopy (CIME) at EPFL for help and insightful discussions. The authors further thank Kuang-Yu Yang from the Nanophotonics and Metrology Laboratory at EPFL for help with the reflection measurements. The authors hereby thank the financial contributors to this work, namely the European Research Council (Starting grant 679211, "Flowtonics") and the Swiss National Science Foundation (Grant 200021_178971).

**Methods**

*Sample Fabrication*

Chalcogenide thin films (Se, $As_2Se_3$) are first thermally evaporated (UNIVEX 350, Oerlikon, Germany) onto three types of substrate: two UV-curable polymers (Ormocomp®, Ormostamp®) well suited for nanoimprint lithography and a pure Silica texture obtained by sol-gel process. We evaporate the films both on textured and non-textured regions to later compare these two relative situations. The film thickness is monitored during evaporation using a quartz crystal (Inficon, Switzerland). Film viscosity is dramatically reduced upon annealing above their glass transition temperature, and enhanced chain mobility allows for dewetting to occur.

*Contact Angle Measurements*

To experimentally determine contact angles, we proceed to dewet evaporated thin films. This is achieved by thermal annealing above the glass transition temperature of the film for extended durations, e.g. twice the time required to observe stable microstructure based on top-view observations using optical microscopy. Cross-sections of the obtained samples are then prepared using liquid nitrogen. All SEM samples were coated with a 10 nm carbon film. The SEM images were taken with a Zeiss Merlin field emission SEM equipped with a GEMINI II column operating at 1.0 kV with a probe current of 70 pA. Contact angles are further measured by Scanning Electron

Microscopy cross-sectional images and processed using image analysis to accurately extract the contact angles (Image J, Contact Angle Module).

*Numerical Simulations*

The numerical implementation of the lubrication equation (1) with complete curvature (5), and together with the interface potential expressions (2), is performed in the finite-element solver COMSOL Multiphysics. The equations are discretized for the variables $(h, \kappa)$. We consider quadratic Lagrangian elements for the spatial discretization, with a triangular non-structured grid for the two-dimensional case. We exploit the built-in Backward Differentiation Formula algorithm for the time marching, setting a tolerance of $10^{-5}$. The numerical convergence is achieved by performing several simulations with $h_s = 0$ and verifying the convergence of the contact angle to the desired value.

As outlined above, the approach for the simulation of experimental conditions is based on the choice of the contact angle and the retrieval of the equilibrium thickness and Born coefficient. The Lennard-Jones potential reads:

$$\Pi = -\frac{\partial \varphi}{\partial h} = \frac{8B}{h^9} - \frac{A}{6\pi h^3} \quad (A1)$$

where the Hamaker constant A is estimated based on the Lifschitz theory. Following reference 24, the macroscopic contact angle at the equilibrium is given by:

$$1 + tan^2 \theta = \left(\frac{\varphi(h_{eq})}{\gamma} + 1\right)^{-2} \quad (A2)$$

where φ(h$_{eq}$) is the equilibrium potential, obtained imposing φ'(h$_{eq}$)=0, where h$_{eq}$ is the equilibrium thickness (i.e. precursor film thickness). Once the contact angle is fixed, the previous relation gives a unique value of the equilibrium potential in the range [0°, 90°], with A>0. The value of the equilibrium potential can be used to evaluate the equilibrium thickness. Deriving expression (2) with respect to h and evaluating at the equilibrium thickness $h_{eq}$ yields:

$$0 = -\frac{8B}{h_{eq}^9} + \frac{A}{6h_{eq}^3} \quad (A3)$$

$$h_{eq} = \left(\frac{48\pi B}{A}\right)^{1/6} \tag{A4}$$

The equilibrium potential at the equilibrium thickness can be written as (c.f. equation (2)):

$$B = h_{eq}^8 \varphi(h_{eq}) + \frac{A}{12\pi} h_{eq}^6 \tag{A5}$$

Where $\varphi(h_{eq})$ is associated to a unique contact angle between 0° to 90° according to equation (4). We therefore obtained the Born coefficient that satisfies the imposed contact angle.

*Evaluation Of The Hamaker Constant Using The Lifschitz Theory And Typical Values*

The Hamaker constant (A) quantifies the imbalance in Van Der Waals forces as two interfaces are brought closer to each other. Lifshitz[20] developed a theory to account for the collective interactive forces between macroscopic particles from quantum field theory that relates the interaction energy with the interparticle distance. The interactions between the particles are relative to the macroscopic properties: the dielectric constant, ε, and the refractive index, n. The Hamaker constant of a system made of a liquid film (3) placed in between a gas or immiscible liquid (2), and a solid (1) can be estimated by considering the overall system energy, which includes (i) permanent polar dipole interactions (Keesom and Debye molecular forces) and (ii) induced dipole interactions (London dispersion forces), which depend on orbiting electron frequency, ν, and the refractive index, n, of the media:[23]

$$A = A_{\nu=0} + A_{\nu>0}$$

$$A \approx \frac{3kT}{4}\left(\frac{\varepsilon_1(0) - \varepsilon_3(0)}{\varepsilon_1(0) + \varepsilon_3(0)}\right)\left(\frac{\varepsilon_2(0) - \varepsilon_3(0)}{\varepsilon_2(0) + \varepsilon_3(0)}\right) + \frac{3h\nu_e}{8\sqrt{2}} \frac{(n_1^2 - n_3^2)(n_2^2 - n_3^2)}{(n_1^2 + n_3^2)^{\frac{1}{2}}(n_2^2 + n_3^2)^{\frac{1}{2}}\left((n_1^2 + n_3^2)^{\frac{1}{2}} + (n_2^2 + n_3^2)^{\frac{1}{2}}\right)}$$

Where $\nu_e$ is the principal UV absorption frequency (~ $3.10^{15}$ Hz), $n_i$ refers to the visible real refractive index of specie i, and $h$ is the Planck constant. Unless strongly polar molecules are involved, the first term can be safely neglected.[23]

Refractive indices in the visible

TABLE 1. Refractive indices in the visible of the typical materials involved in the present study.

| Material | Refractive index | Details | Reference |
|---|---|---|---|
| $SiO_2$ | 1.45 | @550nm, fused silica | 35 |
| As2Se3 | 3.5 | @550nm | Ellipsometry Measurement |
| Ormocomp (OC) | 1.52 | @589nm | From Fabricant |

Based on the Lifschitz theory and the refractive indices provided above, the following Hamaker constants can be evaluated:

$$A_{OC-As_2Se_3-Air} = 5.7 \cdot 10^{-19} J$$

$$A_{SiO_2-As_2Se_3-Air} = 5.9 \cdot 10^{-19} J$$

*Roughness Measurements*

Atomic force microscopy images (Figure S.I. 6) were collected in amplitude modulation mode on a commercial Cypher S system (Asylum Research/Oxford Instruments, Santa Barbara, CA). Two kinds of cantilevers were used: the sensitivity of Asyelec cantilevers (Asylum Research) was evaluated from force curves and the spring constant was measured from their thermal spectra, while AC240TS cantilevers (Asylum Research) were calibrated using the built-in GetReal Automated Probe Calibration procedure. The cantilevers were driven acoustically. Using Gwyddion post-processing software, polynomial plane leveling of order 2 was achieved followed by scar removal using the in-built functions.

*Reflection Measurements*

Reflection spectra were characterized using the Nikon Optiphot 200 inspection microscope (10×, NA = 0.25 objective). A CCD camera (Digital Sight DS-2Mv, Nikon) was used to record the images of the sample, and the images were processed with NIS-Elements F3.2 software. The spectra were characterized with a visible-nIR spectroscopy system based on an inverted optical microscope (Olympus IX-71) coupled to a spectrometer (Jobin Yvon Horiba Triax 550). The sample was illuminated using a halogen white light source focused onto the sample using an objective (20×, NA = 0.4). The reflected light was collected through the same objective and recorded using a spectrometer. The reflected intensity was normalized by the spectrum of the lamp obtained by reflection measurements with a silver mirror (Thorlabs PF 10-03-P01). A polarizer (WP25M-UB, Thorlabs) was used to set linear polarized light illumination for both reflection and transmission measurements.

# Supplementary Information

**Prediction of self-assembled dewetted nanostructures for photonics applications via a continuum mechanics framework**

L.Martin-Monier[1,*], P.G. Ledda[2,*], P.L Piveteau[1], F. Gallaire[2,†], F. Sorin[1,†]

[1] Laboratory of Photonic Materials and Fiber Devices, École Polytechnique Fédérale de Lausanne, 1015, Lausanne, Switzerland.

[2] Laboratory of Fluid Mechanics and Instabilities, École Polytechnique Fédérale de Lausanne, 1015, Lausanne, Switzerland.

*: These authors contributed equally to this work

†: Corresponding Author

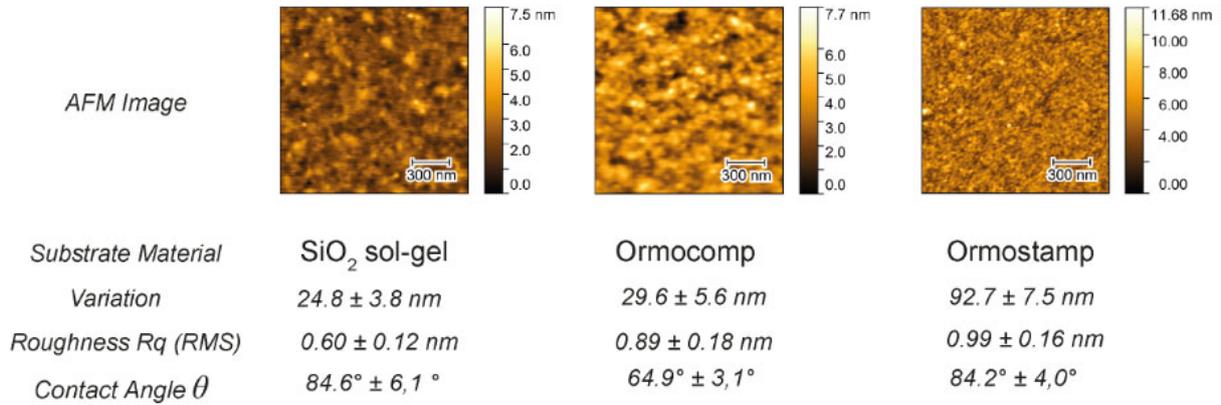

FIG. SI 1 – Roughness parameters of the substrates used in his work. Root mean square Roughness exhibits values <1 nm, while the variation. Variation values (i.e. integral of the absolute value associated with the local gradient) show denser features in Ormostamp substrates than in other elements.

To limit contact angle hysteresis, we proceed in successive steps. The first step consists of a preparation of substrates with low surface roughness. We resort to several substrates, including: (i) sol-gels based on acid-catalyzed Methy (triethoxysilane), followed by a pyrolysis step at 400°C, or (ii) UV curable commercial resins, in particular Ormocomp® and Ormostamp® from Microresist, Germany. The pyrolysis step for sol-gels helps to densify the resulting silica structure while removing residual organic components. To control the resulting surface roughness of silica-based on sol-gel processes as well as the commercial UV resins, we proceed to measure by atomic force microscopy the surface roughness (FIG. SI 1). All root mean squared (RMS) roughness values are inferior to 1 nm, which, although not competitive with typical Si wafer roughness, compares favorably with most other surfaces.

A second step aims at cleaning thoroughly the substrates to remove any chemical inhomogeneities. As reported in the literature, this is a critical step to ensure a homogeneous substrate surface energy, and consequently homogeneous dewetting patterns. The extensive use of polymers largely restricts the use of commonly-used aggressive solvents to clean wafers such as $H_2SO_4$ or HF. Nevertheless, common cleaning procedures help to wash away nano-imprinting residues such as silicone oil traces from the PDMS or other contaminants, All substrates are subsequently washed systematically using isopropanol (degreasing agent), ethanol, and water. The wash cycle is commonly repeated three times, followed by a gentle nitrogen or air gun to dry the substrate surface and blow away eventual debris or particles remaining at the surface.

The hysteresis caused by roughness and chemical heterogeneity is further assessed by evaluating the standard deviation of contact angles measurement for a given film/substrate couple. The error bar stem from a combination of imaging analysis related uncertainties and substrate surface roughness, which induce local triple line pinning and deviation for equilibrium contact angle. As apparent in FIG. SI 1, limited contact angle hysteresis does occur in all systems studied. Further techniques to decrease the surface roughness of nanoimprinted substrate as well as improve the chemical washing procedure would be of considerable help to reduce contact angle standard deviation and thereby offer better control over templated dewetting.

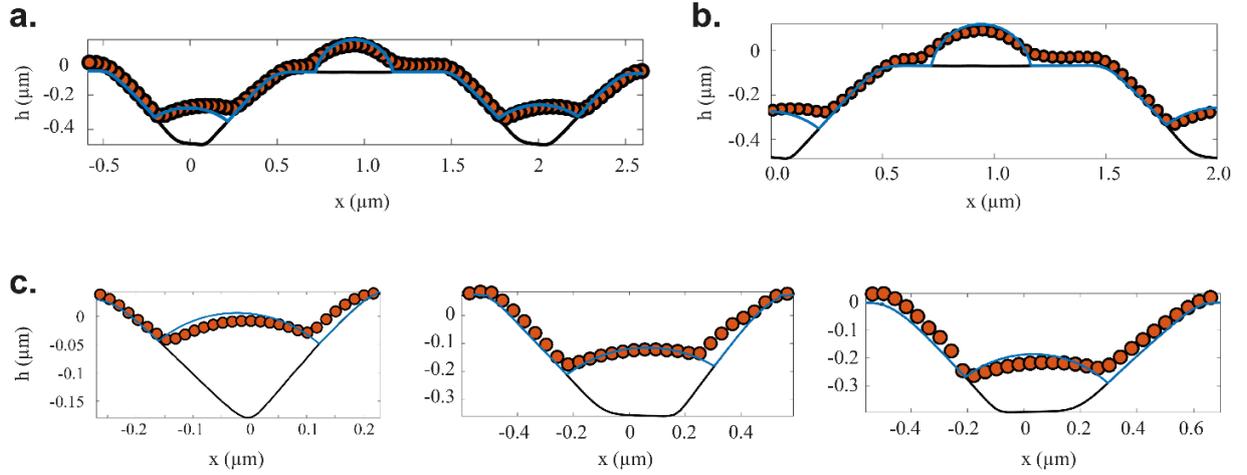

FIG. SI 2 – Additional comparative analysis between experimental results (orange dots) and simulations (solid blue line) for various 2D templates with varying spacing. Experimental data are obtained using $As_2Se_3$ thin films deposited onto plasma-treated Ormocomp substrates. (a)-(b) Final film profile for a 1D textured substrate with pyramid width W = 850 nm and inter-pyramid spacing S = P-W = 1 µm. (b) represents a zoom over a single unit period of (a), highlighting the reasonable overlap between simulated and experimental height profiles. The initial deposited film thickness is $h_i$=60 nm. (c) Final film profile for a 1D textured substrate with pyramid width W = 850 nm and inter-pyramid spacing S = P-W = 100 nm. The initial film thickness is $h_i$=60 nm.

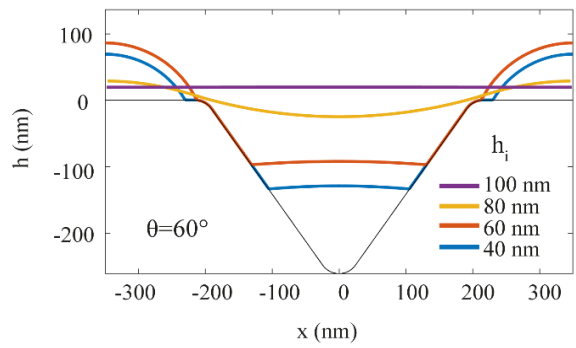

FIG. SI 3 – Plot representing the final thickness profile at the end of simulation for films of initial increasing thicknesses $h_i$. For $h_i$=60 nm and $h_i$=40 nm, the final profile shows two separated droplets. For $h_i$ = 100 nm and $h_i$ = 80 nm, the final film profile is continuous, indicating that the film will break up according to nucleation and growth of holes at random instead of prescribed locations. The critical thickness hence verifies 60 nm < $h_{crit}$ < 80 nm.

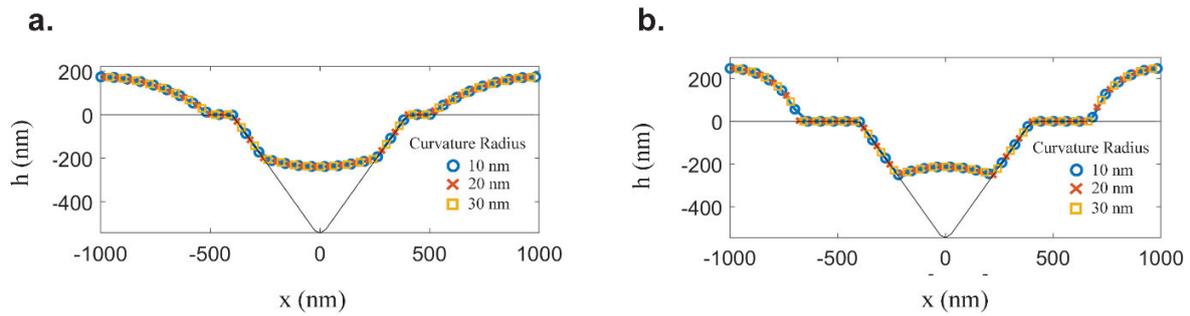

FIG. SI 4 – Influence of the curvature radius of the dewetted structures in 2 dimensions. Each plot reproduces the final film profile for three different edge curvatures: 10 nm (blue), 20 nm (orange), 30 nm (yellow). All final profiles overlap closely and are indistinguishable. (a) Contact Angle 40° and (b) Contact angle 80°. The initial film height is 100 nm

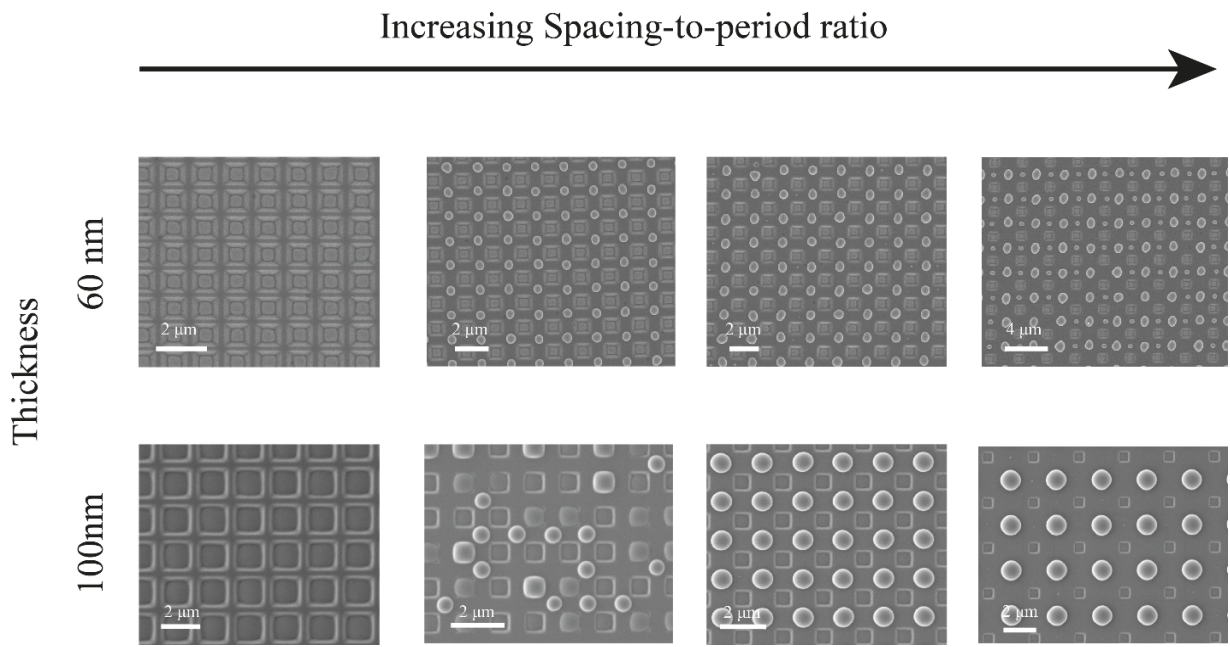

FIG. SI 5 – Top view scanning electron microscopy (SEM) images of an As2Se3 film with increasing thickness deposited on a plasma-treated silica textured sample with increasing spacing-to-period ratio. The structural transition from the absence of mesa (e.g. particle in between pits) to single mesa and then double mesa is visible with an initial film thickness of 60 nm. In the 100 nm thickness case, the film only shows a single structural transition from the absence of mesa to single mesa.

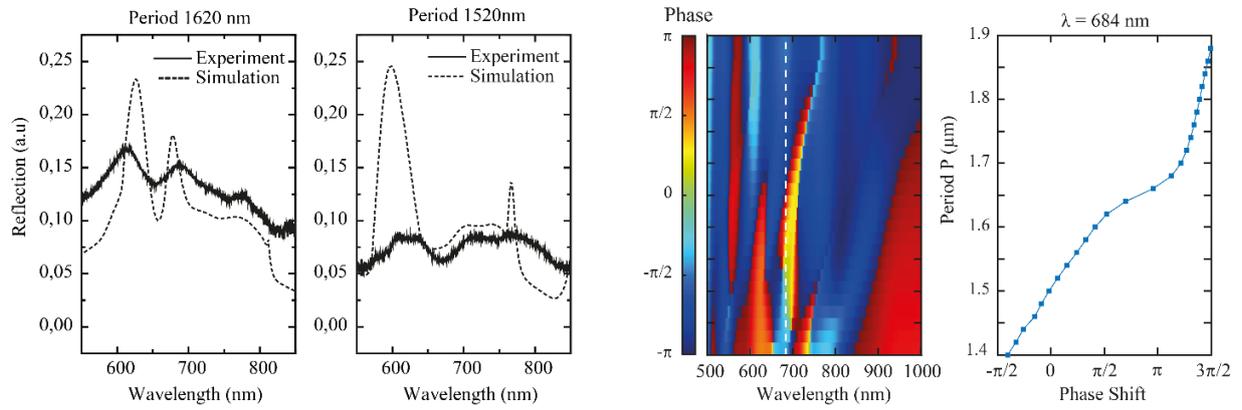

FIG. SI 6 – All-dielectric Huygens Meta-Gratings for linear features according to the geometry shown in FIG. 3(a). The simulated dash curves are modeled purely enforcing the volume conservation hypothesis, with respective periods of 1620 nm and 1520 nm and an initial film thickness of 85 nm. The experimental data is obtained for a film thickness of 80 nm based on microbalance measurements in-situ during evaporation and respective periods of 1600 nm and 1500 nm. The colormap shows the phase map in reflection, with a sharp transition around 690 nm. The phase shift is plotted for λ=684 nm, showing a clear 2π phase shift for a range of periods spanning from 1400 nm to 1900 nm.